\documentclass[pre,aps,twocolumn]{revtex4-2}
\usepackage[T1]{fontenc}
\usepackage[utf8]{inputenc}
\usepackage{amsmath}
\usepackage{amsfonts}
\usepackage{amssymb}
\usepackage{mathtools}
\usepackage{graphicx}
\usepackage[english]{babel}
\usepackage{xcolor}
\usepackage[hidelinks]{hyperref}
\usepackage[capitalise]{cleveref}
\usepackage{siunitx}
\usepackage{blindtext}

\newcommand{\su}[1]{^\mathrm{#1}}
\newcommand{\ind}[1]{_\mathrm{#1}}
\newcommand{\Fid}[1]{F\su{id}[#1]}
\newcommand{\Fex}[1]{F\su{exc}[#1]}
\newcommand*\diff{\mathop{}\!\mathrm{d}}
\renewcommand*{\vec}[1]{\boldsymbol{#1}}
\newcommand*{\intd}[1]{\int\!\diff #1\,}
\newcommand*{\intdz}[0]{\int_{0}^{h}\diff z \,}
\newcommand{\funcd}[2]{\frac{\delta #1}{\delta #2}}
\newcommand{\rhom}[0]{\rho_m}

\newcommand{\mum}[0]{\mu_m}
\newcommand{\kB}[0]{k\ind{B}}
\newcommand{\kT}[0]{\kB T}
\newcommand{\ecp}[0]{\eta_{\rm cp}}
\newcommand{\mue}[0]{\mu_{\rm eff}}
\newcommand{\fp}[0]{f_{\rm P}}
\newcommand{\rhoe}[0]{\rho_{\rm EOS}}
\newcommand{\etae}[0]{\eta_{\rm EOS}}
\newcommand{\etacp}[0]{\eta_{\rm cp}}
\newcommand{\mb}[0]{\bar{m}}
\newcommand{\mbuoy}[0]{m_{b}}
\newcommand{\etab}[0]{\bar{\eta}}
\newcommand{\rhobm}[0]{\bar{\rho}_m }

\newcommand{\eg}{e.g.\ }
\newcommand{\ie}{i.e.\ }
\newcommand{\wrt}{w.r.t.\ }

\newcommand{\crefsub}[3]{Figs.~\ref{#1}(#2) and \ref{#1}(#3)}

\newcommand{\citeau}[1]{\citeauthor{#1}~\cite{#1}}

\graphicspath{{./}}

\begin{document}

\author{Tobias Eckert}
\affiliation{Theoretische Physik II, Physikalisches Institut,
  Universit{\"a}t Bayreuth, D-95440 Bayreuth, Germany}

\author{Matthias Schmidt}
\email{Matthias.Schmidt@uni-bayreuth.de}
\affiliation{Theoretische Physik II, Physikalisches Institut,
  Universit{\"a}t Bayreuth, D-95440 Bayreuth, Germany}

\author{Daniel de las Heras}
\email{delasheras.daniel@gmail.com}
\homepage{www.danieldelasheras.com}
\affiliation{Theoretische Physik II, Physikalisches Institut,
  Universit{\"a}t Bayreuth, D-95440 Bayreuth, Germany}

\title{Sedimentation path theory for mass-polydisperse colloidal systems}

\date{\today}

\begin{abstract}
	Both polydispersity and the presence of a gravitational field are inherent to essentially any colloidal experiment.
	While several theoretical works have focused on the effect of polydispersity on the bulk phase behavior of a colloidal system, little is known about the effect of a gravitational field on a polydisperse colloidal suspension.
	We extend here sedimentation path theory to study sedimentation-diffusion-equilibrium of a mass-polydisperse colloidal system: the particles possess different buoyant masses but they are otherwise identical.
	The model helps to understand the interplay between gravity and polydispersity on sedimentation experiments.
	Since the theory can be applied to any parent distribution of buoyant masses, it can be also used to study sedimentation of monodisperse colloidal systems.
	We find that mass-polydispersity has a strong influence in colloidal systems near density matching for which the bare density of the colloidal particles equals the solvent density.
	To illustrate the theory, we study crystallization in sedimentation-diffusion-equilibrium of a suspension of mass-polydisperse hard spheres.
\end{abstract}

\maketitle

\section{Introduction}

A certain degree of polydispersity in \eg the size and the shape of the particles is inherent to all natural colloids.
Even tough modern synthesis techniques allow the preparation of almost monodisperse colloidal particles~\cite{Murphy2017,Roller2020,Voggenreiter2020,Roller2021}, a small degree of polydispersity is unavoidable.
Understanding bulk phase equilibria in polydisperse systems is a significant challenge~\cite{Sollich2001}.
Polydispersity alters the relative stability between bulk phases~\cite{Kooij2000b,MartinezRaton2002,Fasolo2005,doi:10.1073/pnas.1909357117,PhysRevE.106.034602}.
Phases that are metastable in the corresponding monodisperse system can become stable due to polydispersity.
Examples are the occurrence of hexatic columnar~\cite{PhysRevLett.95.077801} and smectic phases~\cite{Sun2009} in polydisperse discotic liquid crystals, as well as macrophase separation in diblock copolymer melts~\cite{PhysRevLett.99.148304}.
The opposite phenomenon can also occur.
For example, crystallization in a suspension of hard-spheres is suppressed above a terminal polydispersity~\cite{Pusey1987,PhysRevE.59.618,Auer2001}.
Also, fractionation into several phases appears if the degree of polydispersity is high enough~\cite{PhysRevLett.91.068301,PhysRevLett.104.118302,Sollich2011}.
A smectic phase of colloidal rods is no longer stable above a terminal polydispersity in the length of the particles~\cite{Bates1998}.
Dynamical processes such as shear-induced crystallization~\cite{Masshoff2020} are also affected by polydispersity.
During drying, a strong stratification occurs in polydisperse colloidal suspensions~\cite{Fortini2017,Cusola2018}, and the dynamics of large and small particles is different if the colloidal concentration is large enough~\cite{Zaccarelli2015,PhysRevLett.119.048003}.

Sedimentation-diffusion-equilibrium experiments are a primary tool to investigate bulk phenomena in colloidal suspensions.
However, the effect of the gravitational field on the suspension is far from trivial~\cite{Kooij2001a,Piazza2012a,Heras2012,Piazza2014,Pariente2022} and it needs to be understood in order to draw correct conclusions about the bulk~\cite{Eckert2021}.
Gravity adds another level of complexity to the already intricate bulk phenomena of a polydisperse suspension.
To understand the interplay between sedimentation and polydispersity, we introduce here a mass-polydisperse colloidal suspension:
a collection of colloidal particles with the same size and shape (and also identical interparticle interactions) but with buoyant masses that follow a continuous distribution.
Since the interparticle interactions are identical, mass-polydispersity does not have any effect in the bulk phase behaviour.
Hence, our model isolates the effects of a gravitational field on a polydisperse colloidal system from the effects that shape- and size-polydispersity generate in bulk. 

We formulate a theory for mass-polydisperse colloidal systems in sedimentation-diffusion-equilibrium. The theory is based on sedimentation path theory~\cite{Heras2013,Geigenfeind2016} which incorporates the effect of gravity on top of the bulk description of the system.
Sedimentation path theory uses a local equilibrium approximation to describe how the chemical potential of a sample under gravity changes with the altitude.
So far, sedimentation path theory has been used to study sedimentation in colloidal binary mixtures~\cite{Heras2012,Heras2013,Drwenski2016,PhysRevE.93.030601,Geigenfeind2016,Avvisati2017,Dasgupta2018,BrazTeixeira2021,Eckert2021,PhysRevResearch.4.013189}.
In this work, we extend sedimentation path theory to mass-polydisperse systems.
Using statistical mechanics, we obtain the exact expression for the sedimentation path of the mass-polydisperse suspension combining the individual paths of all particles in the distribution.
We use a model bulk system to illustrate and highlight the key concepts of the theory, such as the construction of the sedimentation path and that of the stacking diagram (which is the analogue of the bulk phase diagram in sedimentation).
The theory is general and can be applied to any colloidal system in sedimentation-diffusion-equilibrium.
Moreover, the theory contains the description of a monodisperse system as a special limit (delta distribution of the buoyant masses).
As a proof of concept, we study sedimentation of a suspension of mass-polydisperse hard-spheres with different buoyant mass distributions.
We find that mass polydispersity plays a major role in systems near density matching.
For example, near density matching the packing fraction and the height of the sample at which crystallization is observed in sedimentation-diffusion-equilibrium are strongly influenced by the details of the mass distribution.

\section{Theory}

\subsection{Bulk}
\label{sec:bulk}

We use classical statistical mechanics to describe the thermodynamic bulk equilibrium of our mass-polydisperse colloidal system.
The term bulk refers here to an infinitely large system in which boundary effects can be neglected and that is not subject to any external field.
The particles differ only in their buoyant masses.
Since the buoyant mass does not play any role in bulk, the bulk phenomenology of our model is identical to that of a monocomponent system in which only one buoyant mass is present.
Only when gravity is incorporated into both systems the buoyant mass becomes a relevant parameter and the behaviour of the mass-polydisperse and the monodisperse colloidal systems will differ from each other.

The total Helmholtz free energy $F$ is the sum of the ideal and the excess contributions, \ie $F=F\su{id}+F\su{exc}$.
In a mass-polydisperse system, the free energy is a functional of $\rhom$, the density distribution of species with buoyant mass $m$.
For simplicity, we work with a scaled, dimensionless, buoyant mass $m=\mbuoy/m_0$, where $\mbuoy$ is the actual buoyant mass of a particle and $m_0$ is a reference buoyant mass.
Sensible choices relate $m_0$ to \eg the average buoyant mass of the distribution or its standard deviation.
The concrete definition of $m_0$ is given below in each considered system.

The ideal contribution to the free energy is a functional of $\rhom$ and is given exactly by
\begin{equation}
  \label{eq:Fid}
  \Fid{\rhom} = \kT \intd{m} \rhom (\ln(\rhom) - 1),
\end{equation}
where $\kB$ is the Boltzmann's constant and $T$ is the absolute temperature.
Without loss of generality we measure $\rhom$ relative to the thermal de Broglie wavelengths $\Lambda_m = \sqrt{2\pi \hbar ^{2}/(\mbuoy\kT)}$ with reduced Planck's constant $\hbar$.
Note that the value of $\Lambda_m$ does not play any role here since altering $\Lambda_m$ simply adds a term to the free energy that is proportional to the total number of particles with buoyant mass $m$. Such term can be reinterpreted as a change of the origin of the chemical potential of the species with buoyant mass $m$.

The integration over $m$ in~\cref{eq:Fid} reflects the fact that due to the mass-polydispersity, the buoyant mass is a continuous variable.
For the shake of simplicity, we omit the positional argument $\vec{r}$ in the density distribution as well as its corresponding space integral that appear in bulk-phases with positional order such as crystalline phases.

The ideal free energy, Eq.~\eqref{eq:Fid}, accounts for the entropy of mixing of our mass-polydisperse system.
The overall density across all species $\rho$ follows directly from the density distribution of buoyant masses
\begin{equation}
  \label{eq:rho}
  \rho = \intd{m} \rhom.
\end{equation}
Since the interparticle interaction is independent of the buoyant masses of the particles, only the density across all species $\rho$ enters into the excess (over ideal) free energy.
Hence, the excess free energy functional must satisfy
\begin{equation}
  \label{eq:Fex}
  \Fex{\rhom} = \Fex{\rho}.
\end{equation}

The grand potential is also a functional of $\rhom$ given by
\begin{equation}
  \label{eq:Omega}
  \Omega[\rhom] = \Fid{\rhom} + \Fex{\rho} - \intd{m} \rhom \mum,
\end{equation}
where $\mum$ is the chemical potential of the species with buoyant mass $m$.
In equilibrium $\Omega[\rhom]$ is minimal \wrt the mass-density distribution, \ie 
\begin{equation}
  \label{eq:Omega-min}
	\funcd{\Omega[{\rhom}]}{\rho_{m'}} = 0.
\end{equation}

The Euler-Lagrange equation associated to~\cref{eq:Omega-min}, see derivation in Appendix~\ref{sec:A1}, reads
\begin{gather}
  \ln(\rhom) - \ln(\rho) + \beta \mu - \beta \mum = 0,
	\label{eq:EL}
\end{gather}
where $\mu$ is the chemical potential of a monodisperse system with overall density $\rho$, see~\cref{eq:rho}.
Hence, it follows from Eq.~\eqref{eq:EL} that the density of particles with buoyant mass $m$ can be written as
\begin{equation}
  \label{eq:rhom}
\rhom = \rho e^{\beta(\mum - \mu)}.
\end{equation}
Integrating \cref{eq:rhom} over $m$ on both sides, and using~\cref{eq:rho} on the left hand side, leads to
\begin{align}
  \label{eq:mu-int-1}
  \rho = \rho \intd{m} e^{\beta(\mum - \mu)}.
\end{align}
Since $\rho\neq0$, we obtain
\begin{align}
  \label{eq:mu-int}
  e^{\beta\mu} = \intd{m} e^{\beta\mum},
\end{align}
which constitutes an exact analytic expression for the chemical potential of the monodisperse bulk system
\begin{equation}
  \label{eq:mu}
  \mu = \kT \ln \left( \intd{m} e^{\beta \mum} \right),
\end{equation}
in terms of the chemical potentials of the individual species $\mum$ in the mass-polydisperse system.
In a monodisperse system there exists only a single species and Eq.~\eqref{eq:mu} holds trivially.

\subsection{Particle Model}
\label{sec:model}

\begin{figure}
  \centering
  \includegraphics[width=\linewidth]{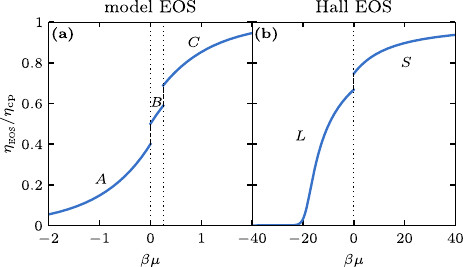}
	\caption{Packing fraction $\etae$ relative to close packing $\ecp$ as a function of the scaled chemical potential $\beta\mu$ for (a) our model equation of state, and (b) the Hall equation of state~\cite{Hall1972} for hard spheres.
	Our model EOS (a) contains three different bulk phases named $A$, $B$ and $C$ which could correspond to \eg the isotropic, the nematic, and the smectic phases of a lyotropic liquid crystal.
	The Hall EOS (b) describes the liquid ($L$) and the solid crystalline ($S$) phases of a hard-sphere system.
	The vertical dotted lines indicate the chemical potentials of the different bulk phase transitions.
	Without loss of generality, we have translated the origin of chemical potential such that it coincides with the chemical potential of (a) the $A$-$B$, and (b) the $L$-$S$ transitions.
  }
  \label{fig1}
\end{figure}

To proceed we need the bulk equation of state (EOS) of the monodisperse colloidal system, $\rhoe(\mu)$.
Given an interparticle interaction potential, several methods can be used to obtain the corresponding bulk EOS.
These include \eg density functional theory~\cite{Evans1979}, liquid state integral equation theory~\cite{Chiew1990,Schweizer1988,Lado1968}, computer simulations~\cite{Johnson1993,SaikaVoivod2000,Rowley1997} and empirical expressions~\cite{Span1996,Soave1972,Peng1976}.
Here, and with the only purpose of illustrating our theory we use a model (fabricated) EOS that contains two phase transitions, see~\cref{fig1}(a).
Our model EOS satisfies both the ideal gas limit
\begin{equation}
  \lim_{\mu \rightarrow -\infty}\rhoe(\mu) \sim e^{\beta\mu},
\end{equation}
and also the close packing limit characteristic of systems with hard core interactions
\begin{equation}
  \lim_{\mu \rightarrow \infty}\etae(\mu)/\etacp = 1,
\end{equation}
here $\etae$ is the packing fraction (percentage of volume occupied by the particles) according to the EOS and $\etacp$ is the close packing fraction.
Such EOS could represent \eg a lyotropic colloidal system with two first-order bulk phase transition, say isotropic-nematic and nematic-smectic.

Apart from the model EOS, we also illustrate and validate the theory by studying sedimentation of a suspension of hard-spheres.
We use the analytical EOS proposed by~\citeau{Hall1972}, which describes the liquid ($L$) and solid crystalline ($S$) phases of a hard sphere system.
The Hall EOS was originally formulated using the compressibility factor as a function of the density.
Following Ref.~\cite{Mulero1999}, we numerically integrate the analytical Hall EOS to obtain the chemical potential as a function of the density, see~\cref{fig1}(b) for a graphical representation.
It is sufficient to fix $\rhoe(\mu)$ up to an arbitrary additive constant in $\mu$.
Hence, for convenience, we choose $\mu=0$ as the chemical potential at the liquid-solid first order phase transition.

\subsection{Sedimentation}

\begin{figure*}
  \centering
  \includegraphics[width=\linewidth]{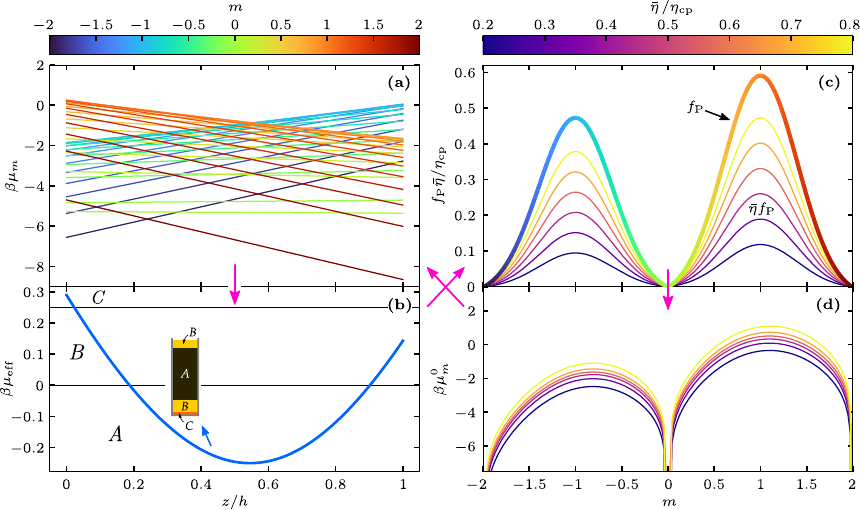}
	\caption{(a) Local chemical potentials $\beta\mu_m(z)$ as a function of the elevation $z$ scaled with the sample height $h$.
	Each sedimentation path varies linearly with $z$ and it is colored according to the buoyant mass of the species $m$ (top-left color bar).
	(b) Effective chemical potential $\beta\mue(z)$ as a function of the scaled elevation $z/h$.
	The non-linear sedimentation path of the mass-polydisperse system (b) is the result of combining the sedimentation paths of each species (a) according to the \texttt{LogSumExp} structure in \cref{eq:mu-eff}.
	The sedimentation paths are defined in the interval $0\leq z\leq h$, and the sample height is $h=2\xi$.
	(c) Imposed parent distribution $\fp(m)$, and plots of the parent distribution scaled with several values of the average packing fraction $\etab$ (see color bar).
	The imposed $\fp(m)$ and the value of $\etab$ in (c) fix the offsets  $\beta\mum^0$ for all buoyant masses $m$, which are shown in panel (d).
	The pink arrows illustrate the iterative procedure to find the effective sedimentation path: for a fixed distribution $\fp(m)$ and packing fraction $\etab$ (c), we give an initial guess for the offsets $\mum^0$ (d), calculate the individual paths $\mu_m(z)$ (a), and combine them to get the effective path $\mue(z)$ (b). Using the effective path we obtain the density profile and then the resulting distribution of particles and the average packing fraction. With this information, we readjust the offsets $\mum^0$ until the output distribution and packing fraction are the desired ones.
  }
  \label{fig2}
\end{figure*}

To incorporate gravity into our theory, we extend sedimentation path theory~\cite{Heras2013,Heras2015} as formulated for finite height samples~\cite{Geigenfeind2016,Eckert2021} to include mass-polydispersity.
As often done in colloidal sedimentation, we assume that all horizontal slices of a sample in sedimentation-diffusion-equilibrium can be described as a bulk equilibrium state, and also that they are independent of each other.
This local-equilibrium approximation is justified if the correlation lengths are small compared to the gravitational lengths $\xi_m=\kT/(\mbuoy g)$, which is the case in many colloidal systems.
Here $g$ is the acceleration of gravity.

We work in units of the thermal energy $\kT$, the gravitational constant $g$, and the reference mass $m_0$ for ease of comparability between different systems.
Using $m_0$ we define a reference gravitational length $\xi = \kT/(m_0g)$, which acts as our fundamental length scale.

We treat the slices for each elevation $z$ as a bulk system with local chemical potentials for each species $\mu_m$ given by
\begin{equation}
  \label{eq:mum}
  \mum(z) = \mum^0 - \mbuoy gz,
\end{equation}
Here $\mum^0$ is the chemical potential of the species with buoyant mass $m$ at elevation $z=0$.
The set of constant offsets $\mum^0$ in $\mum(z)$ is \emph{a priori} unknown and must be determined via an iterative numerical procedure to match the prescribed mass-resolved density distribution $\rhom$.
Returning to the discussion about the thermal wavelengths, altering the value of $\Lambda_m$ would only
introduce a constant term $\ln(\Lambda_m)$ in Eq.~\eqref{eq:EL} that can be reabsorbed in Eq.~\eqref{eq:mum} as a shift of the chemical potential $\mu_m$ via the offset $\mum^0$.
The offsets $\mum^0$ depend therefore on the choice of $\Lambda_m$.
However, the sedimentation profiles $\rho_m(z)$ remain unchanged, since $\mum^0$ are determined to match the prescribed density distribution.

Equation~\eqref{eq:mum} is the sedimentation path~\cite{Heras2013,Heras2015,Geigenfeind2016,Eckert2021} of the species with buoyant mass $m$.
It hence describes how the chemical potential of each species varies linearly with $z$ in the range $0\le z\le h$, with $h$ the sample height.
The local chemical potential for each species either decreases ($\mbuoy>0$) or increases ($\mbuoy<0$) with the elevation $z$, depending on the sign of the buoyant mass.

The sedimentation path of each species $\mu_m(z)$ is just a straight line, see~\cref{fig2}(a), as in the case of monodisperse systems. 
Next, we combine all paths at each elevation $z$ to obtain an effective chemical potential $\mue(z)$.
Inserting $\mum(z)$ in~\cref{eq:mum} into~\cref{eq:mu} yields the sedimentation path of a mass-polydisperse system
\begin{equation}
  \label{eq:mu-eff}
  \mue(z) = \kT \ln \left( \intd{m} e^{\beta(\mum^0 - \mbuoy gz)} \right).
\end{equation}
Equation~\eqref{eq:mu-eff}, which has the form of a \texttt{LogSumExp} function, describes how the effective chemical potential of the mass-polydisperse system varies vertically along the sample in sedimentation-diffusion-equilibrium.
We give an example of $\mue(z)$ in ~\cref{fig2}(b).
The sedimentation path is obtained from the set of $\mum(z)$ in~\cref{fig2}(a) via~\cref{eq:mu-eff}.
The sedimentation path is no longer a straight line even though the individual paths for each species are lines.
Since (i) the logarithm is a concave function, (ii) the scalars $\exp(\beta\mum^0)$ are positive, and (iii) the exponential is a convex function, it follows that $\mue(z)$, as given by Eq.~\eqref{eq:mu-eff}, is a convex function of the elevation $z$.
This is a strong constraint on the possible shapes of $\mue(z)$.
It means that (i) $\mue(z)$ can have only one minimum and also that (ii) the local maxima of $\mue(z)$ in the interval $0 \le z \le h$ are either $z=0$, $z=h$, or both of them.
As we discuss below, the extrema of the path $\mue(z)$ are important because they determine the layers of different bulk phases that form in the sample.

Via the equation of state $\rhoe(\mu)$ for the bulk density we then obtain the density profile across all species
\begin{equation}
  \label{eq:rho-z}
  \rho(z) = \rhoe(\mue(z)),
\end{equation}
at elevation $z$ from~\cref{eq:mu-eff}.

The density of species with buoyant mass $m$ at elevation $z$ follows then by inserting~\cref{eq:mum,eq:mu-eff,eq:rho-z} into~\cref{eq:rhom}
\begin{equation}
  \label{eq:rhomz}
	\rhom(z) = \rho(z) e^{\beta\left(\mum^0 - \mbuoy gz - \mue(z)\right)}.
\end{equation}

The average density of particles with buoyant mass $m$ in a sample with height $h$ is then given by
\begin{equation}
  \rhobm = \frac{1}{h} \intdz \rhom(z).
\end{equation}
The value of $\rhobm$ is also the density of particles with buoyant mass $m$ in the initial distribution, \ie before the particles sedimented and equilibrated.

The average packing fraction is 
\begin{equation}
  \label{eq:etabar}
  \etab = \frac{v_0}{h} \intdz \rho(z),
\end{equation}
where $v_0$ is the particle volume. 

The parent distribution, which gives the overall probability of finding a particle with buoyant mass $m$ anywhere in the sample can be obtained as
\begin{equation}
  \label{eq:fp}
	\fp(m) = \frac{\rhobm}{\intd{m} \rhobm} =\frac AN \intdz \rhom(z), 
\end{equation}
with $N=hA\intd{m} \rhobm=A\intd{m} \intdz \rhom(z)$ the total number of particles, and $A$ being the area of a cross section of the sample.
Both $\etab$ and $\fp(m)$ are directly comparable with experimental results, since $\etab$ is the concentration of particles in the stock solution (before sedimentation)
and $\fp(m)$ describes the mass-polydispersity of the particles, normalized by the total concentration.

From the definitions \eqref{eq:etabar} and \eqref{eq:fp} we can get back the average density of specie $m$ via
\begin{equation}
  \rhobm = \frac{\etab}{v_0} \fp(m).
\end{equation}

To obtain the sedimentation-diffusion-equilibrium of a mass-polydisperse colloidal system, we start prescribing the sample height $h$, the average packing fraction of the sample $\etab$, and the parent distribution $\fp(m)$.
An illustrative parent distribution that contains particles with both positive and negative buoyant masses is shown in~\cref{fig2}(c).
These initial conditions are sufficient to find the as yet undetermined offsets on the chemical potential for each species $\mum^0$, see Eq.~\eqref{eq:mum}.
We discretize $\fp(m)$ and then numerically determine $\mum^0$ via a least square algorithm which iteratively solves for the prescribed $\etab\fp(m)$ in a sample of height $h$.
With the offsets $\mum^0$ we calculate the corresponding $\mue(z)$ via~\cref{eq:mu-eff}.
Next we obtain $\rho(z)$ and $\rhom(z)$ via Eqs.~\eqref{eq:rho-z} and~\eqref{eq:rhomz}, respectively.
The profiles $\rho(z)$ and $\rhom(z)$ determine both $\etab$ and $\fp(m)$ via Eqs.~\eqref{eq:etabar} and~\eqref{eq:fp} respectively.
The least square algorithm finds then the offsets that minimize the difference to the prescribed (target) values of $\etab$ and $\fp(m)$.

For example, we show in~\cref{fig2}(d) the offsets ${\mum^0}$ corresponding to the distribution prescribed in~\cref{fig2}(c).
We discretize in $m$, and hence the number of input variables $\rhobm$ and unknown variables $\mum^0$ is the same.
The self-consistency problem of finding $\mum^0$ is therefore well defined.
The set of sedimentation paths $\mum(z)$ in Fig.~\ref{fig2}(a) are obtained with the offsets calculated in~\cref{fig2}(d).
The effective sedimentation path $\mue(z)$, see Fig.~\ref{fig2}(b), of the mass-polydisperse system follows then from the set of paths for each species $\mum(z)$.

The sedimentation path of the mass-polydisperse system determines the stacking sequence, \ie the set of layers of bulk phases that are observed in the sample under gravity.
Every time the path crosses the coexistence chemical potential of a bulk transition, an interface between the coexisting phases appears in the cuvette.
By looking at the crossings between the sedimentation path and the bulk binodals we determine the stacking sequence and the position of the interfaces between stacks. 
For example, the sequence corresponding to the path in Fig.~\ref{fig2}(b) is $BABC$ (from top to bottom of the sample).

{\bf Extended Gibbs phase rule.} Given the convexity properties of the sedimentation path, recall our discussion below Eq.~\eqref{eq:mu-eff},
we conclude that the maximum number of layers that can appear in a sedimented sample of a mass-polydisperse system is	$2n_b-1$, with $n_b$ the number of different stable phases in bulk.
This corresponds to the stacking sequence of a mass-polydisperse suspension with positive and negative buoyant masses in which all phases occur repeatedly except the middle layer, which corresponds to the bulk phase stable at low chemical potential.
In our model EOS, the stacking sequence with the maximum number of layers is $CBABC$, for which the sedimentation path is similar to the one in Fig.~\ref{fig2}(b) but extended such that it reenters the $C$ region at high elevations.

If the parent distribution contains only buoyant masses of the same sign, then the maximum number of layers in a stacking sequence is simply $n_b$, the number of stable bulk phases.

\subsection{Stacking diagram}

\begin{figure}
  \centering
 \includegraphics[width=\linewidth]{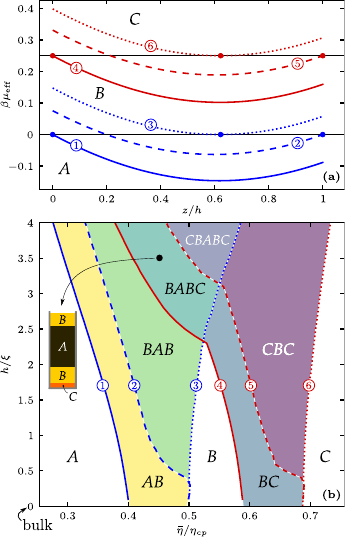} 
	\caption{(a) Sedimentation paths in the plane of effective chemical potential $\beta\mue$ as a function of the elevation $z/h$ for samples of height $h=1.7\xi$ and a parent distribution like in Fig.~\ref{fig2}(c).
	The coexistence chemical potentials for the $A$-$B$ ($\mu\ind{coex}=0$) and for the $B$-$C$ ($\beta\mu\ind{coex}=0.25$) bulk transitions are indicated by solid-black horizontal lines.
	There are six sedimentation paths labeled from $1$ to $6$.
	The average packing fraction $\etab$ of each sample is such that the corresponding path either ends at (solid paths 1 and 4), starts at (dashed paths 2 and 5) or is tangent to (dotted paths 3 and 6) a bulk binodal (horizontal lines). 
   	The points where the paths touch the coexistence bulk chemical potential are marked by solid circles.
	(b) Stacking diagram in the plane of average packing fraction $\etab/\etacp$ and sample height $h/\xi$ for the model EOS in \cref{fig1}(a) and parent distribution as in \cref{fig2}(c).
	The position of the six boundary sedimentation paths in (a) is marked in (b) using the corresponding labels $1$ to $6$.
	The sedimentation binodals of paths that end, start, and are tangent to the bulk binodals are indicated with solid-, dashed- and dotted-lines, respectively.
   	The stacking sequences are labeled from the top of the sample to the bottom.
	Each point in the stacking diagram is a sample in sedimentation. 
	The sketch shows the stacking sequence $BABC$ and relative layer thicknesses of the sample with $\etab/\etacp=0.45$ and $h/\xi=3.5$ (indicated by a black circle).
 }
  \label{fig3}
\end{figure}

Different sedimentation paths can give rise to distinct stacking sequences.
The set of all possible stacking sequences can be represented in a stacking diagram. 
In binary mixtures, the sedimentation paths of both species vary linearly with $z$.
In mass polydisperse systems, we average the linear local chemical potentials $\mu_m(z)$,~\cref{eq:mum}, of all species together, according to~\cref{eq:mu-eff}, and obtain a non-linear effective chemical potential $\mue(z)$.
Even though the sedimentation paths are no longer straight lines, the same ideas as in the case of binary mixtures~\cite{Heras2013,Eckert2021} apply for the construction of the stacking diagram.
In short, we must find all the sedimentation paths that constitute a boundary between two or more stacking sequences in the stacking diagram.
Examples of such paths are shown in Fig.~\ref{fig3}(a).
The boundary paths are the sedimentation paths $\mue(z)$ that either end [paths 1 and 4 in Fig.~\ref{fig3}(a)], start (paths 2 and 5), or are tangent (paths 3 and 6) to a bulk binodal.
These paths are a boundary between two or more stacking sequences since an infinitesimal change of the path in general alters the stacking sequence. 
Without gravity (\ie in bulk) the mass-polydisperse system behaves like a monocomponent system, since the interparticle interaction potential is independent of the buoyant mass.
Thus, in bulk, there is only a single relevant chemical potential.
In the chemical potential vs.\ height plane, the bulk transitions are simply horizontal lines independent of $z$, see~\cref{fig3}(a).
Hence, given that the sedimentation path is convex, a path tangent to a bulk binodal is also a path for which the minimum coincides with the chemical potential of the bulk transition, like \eg paths $3$ and $6$ in Fig.~\ref{fig3}(a). For other types of bulk phase coexistence such as critical and triple points, the procedure to find the boundary paths is the same as the one just described for a bulk binodal.

Next we find the total density profile $\rho(z)$ and the average packing fraction $\etab$ corresponding to each of the boundary sedimentation paths via Eqs.~\eqref{eq:rho-z} and~\eqref{eq:etabar}, respectively.
To obtain the full stacking diagram, we repeat the procedure for every sample height $h$ ranging from zero to the desired maximal sample height.
This provide us with the stacking diagram in the (experimentally relevant) plane of average packing fraction $\etab$ and sample height $h$, see \cref{fig3}(b).
Each point in the stacking diagram represents one sedimentation path and it hence represents one specific sample in sedimentation-diffusion-equilibrium.

For each bulk phase transition there can be at most three boundary lines in the stacking diagram, so-called sedimentation binodals~\cite{Heras2013,Geigenfeind2016,Eckert2021}.
The sedimentation binodals corresponding to the paths that either start or end at the binodal are always present independently of the parent distribution and the sample height.
On the other hand, the sedimentation binodal corresponding to paths tangent to the bulk transition appears if and only if the sedimentation path presents a minimum at intermediate values of $z$.
It follows from~\cref{eq:mu-eff} that a minimum in $\mue(z)$ not located at the bottom ($z\neq0$) or the top ($z\neq h$) of the sample can appear only if the parent distribution contains both positive and negative buoyant masses. 
Even in that case, there might be sample heights for which the path does not have a minimum at intermediate elevations.

In our illustrative example, there are two bulk phase transition ($A-B$ and $B-C$), see Fig.~\ref{fig1}(a), and the parent distribution is made of particles with positive and negative buoyant masses, see Fig.~\ref{fig2}(c).
The stacking diagram contains six sedimentation binodals, see \cref{fig3}(b).
For sample heights $h/\xi\lesssim0.4$ only two types of sedimentation binodals can be observed.
In this low height regime, we cannot find sedimentation paths tangent to the binodal since $\mue(z)$ does not have a minimum at intermediate elevations $0 < z < h$.

Within our local equilibrium approximation, in the limit $h\rightarrow0$ the sedimentation path reduces to a point and hence the stacking diagram reduces to the bulk phase diagram.
In a real system, confinement and surface effects such as wetting and layering will become relevant in the limit of short sample heights.

{\bf Mass-monodisperse system.} Our method to construct the stacking diagram for mass-polydisperse systems contains as a limiting case the monodisperse system.
In a monodisperse system all particles possess the same buoyant mass.
Hence, ~\cref{eq:mu-eff} reduces to
\begin{equation}
\mue(z) = \mum(z) = \mum^0 - \mbuoy gz.
\end{equation}
Thus, as expected, the sedimentation path of a monodisperse system is the segment of a line, linear in $z$.
In the stacking diagram, only the sedimentation binodals of paths that start, \ie $\mue(h) = \mu\ind{coex}$, or end,~\ie $\mue(0) = \mu\ind{coex}$, at the bulk binodal (given by $\mu\ind{coex}$) appear.
The sedimentation path of a monodisperse system can never have a minimum at intermediate elevations.

\section{Results}
\label{sec:results}

We next apply our theory to the arguably best studied colloidal system to date: hard spheres.
We study sedimentation of a mass-polydisperse hard sphere system using
the Hall equation of state~\cite{Hall1972}, represented in the plane of $\mu$ and $\eta$ in Fig.~\ref{fig1}(b), to describe the bulk of the system.

\subsection{Species-resolved probability distributions in mass-polydisperse systems}

\begin{figure*}
  \centering
  \includegraphics[width=\linewidth]{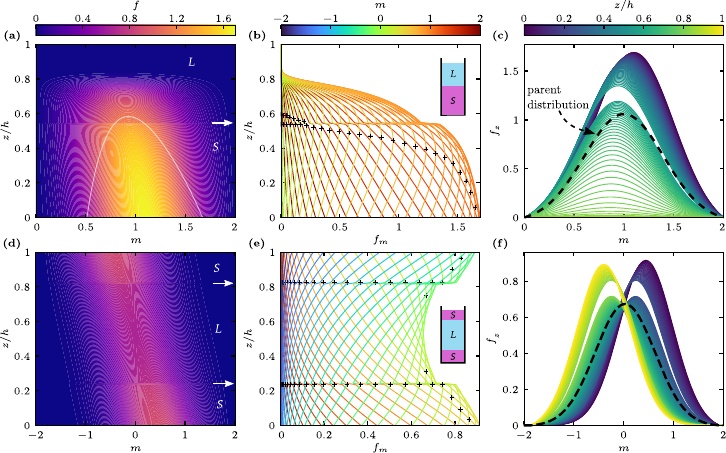} 
  \caption{(a) Probability of finding a particle with buoyant mass $m$ at elevation $z$ relative to the sample height $h$ in a hard sphere system modeled using the Hall EOS.
	The light contour lines in (a) indicate points $(m,z)$  at which the chemical potential $\mum(z)$ [see~\cref{eq:mum}] is equal to the bulk liquid-solid coexistence chemical potential.
	The white arrow indicates the position of the liquid ($L$) - solid crystalline ($S$) interface.
    (b) Vertical slices of panel (a) for fixed buoyant mass $m$ given by the colorbar. The inset is a sketch of the sample.
    (c) Horizontal slices of panel (a) for fixed elevation $z$ given by the colorbar.
	The imposed parent distribution $\fp(m)$ is a Gaussian with standard deviation $0.4$ and centered around $m=1$, \ie the reference buoyant mass $m_0$ is the mean of the parent distribution (see black dashed-line).
    The sample has a height $h=80\xi$ with gravitational length $\xi$ and a packing fraction $\eta=0.6\ecp$, relative to the close packing fraction $\ecp$.
	Panels (d-f) are the same as panels (a-c), but for a Gaussian with standard deviation $0.6$ centered around $0.03$ as the parent distribution $\fp(m)$ (black dashed line), slightly favoring particles with positive buoyant mass, sample height $h=120\xi$ and packing fraction $\eta=0.7\ecp$.
      The crosses in (b) and (e) indicate the position of the local maxima in the probability distribution $f_m(z)$ along elevation $z$ for fixed buoyant mass $m$.
  }
  \label{fig4}
\end{figure*}

The imposed parent distribution of the mass-polydisperse system, $\fp(m)$, describes the probability of finding a particle with a certain buoyant mass $m$ anywhere in the system.
Experimentally, this corresponds to the stock solution.
After letting the dispersion settle under gravity to reach sedimentation-diffusion-equilibrium, a height-dependent density profile develops.
The overall probability distribution integrated over the whole sample is still $\fp(m)$ since particles are conserved.
However, at each horizontal slice the mass composition is generally different from $\fp(m)$.
One expects \eg heavier particles to concentrate next to the bottom of the sample as compared to lighter particles.
Sedimentation path theory allows to carry out a detailed study of the mass distribution along the sample.

We study first a mass-polydisperse dispersion of hard spheres with only positive buoyant mass.
The parent distribution is a Gaussian centered around $m = 1$ and cut at $m=0$ and $m=2$, \ie only buoyant masses in the range $0 \le m \le 2$ are allowed.
The mean packing fraction is $\etab/\etacp=0.6$. Under gravity, the sample develops the stacking sequence: top liquid and bottom solid ($LS$).
We show the probability $f(m,z)$ of finding a particle with buoyant mass $m$ at elevation $z$ in \cref{fig4}(a).
The probability distribution $f_m(z)$ for a fixed buoyant mass $m$ and resolved in $z$, as well as the probability distribution $f_z(m)$ for a fixed $z$ resolved in $m$ are shown in \crefsub{fig4}{b}{c}, respectively.
The distributions $f_m(z)$ and $f_z(m)$ correspond to vertical and horizontal slices of the full distribution $f(m,z)$, respectively.
The distributions $f_z(m)$ are shifted and skewed, Fig.~\ref{fig4}(c), as compared to the parent distribution $\fp(m)$ (black-dashed line) which is symmetric \wrt $m=1$.
As expected, heavier particles are more frequently found at the bottom of the sample.
This becomes more apparent when we look at $f_m(z)$. Fig.~\ref{fig4}(b).
There is a clear depletion of lighter particles from the bottom of the sample.
Interestingly, the probability distribution along $z$ of particles with $m\lesssim1.01$ is not monotonically increasing towards the bottom of the sample, but has a maximum up to $0.5h$ above the bottom.
Lighter particles are displaced by heavier particles from the bottom as a result of a balance between only two contributions: the gravitational energy and the entropy of mixing.
The excess free energy does not play a role in determining the relative position of the particles according to their buoyant masses.
Interchanging heavier for lighter particles and vice versa does not alter the overall density, and thus the excess free energy $\Fex{\rho}$, which is a functional of only the overall density $\rho$, is not affected.

We also show in Figs.~\ref{fig4}(d),~\ref{fig4}(e), and~\ref{fig4}(f), the mass- and height-resolved probability distributions of a sample with a parent distribution containing both positive and negative buoyant masses.
The parent distribution is a Gaussian centered around $m = 0.03$ and cut at $m=\pm1.9$. 
The initial packing fraction is $\etab/\etacp=0.7$ and the stacking sequence is $SLS$.
The liquid-solid interfaces occur at elevations $z/h=0.25$ and $0.8$ and are visible as discontinuities of the distribution functions.
On the top (bottom) of the sample particles with negative (positive) buoyant masses are more frequently found.
This is visible in Fig.~\ref{fig4}(f) as a shift toward negative or positive buoyant masses of the distributions belonging to the solid crystalline layers.

\subsection{Mass-polydispersity close to density matching}
In density matching colloidal experiments, the mass density of the colloidal particles is very close to the mass density of the solvent.
If the density match between particle and solvent is perfect, the buoyant mass of the colloids vanishes and therefore gravity has no effect on the sample.
This, in principle, would allow to carry out a direct comparison between bulk phenomena and sedimentation experiments. 
In practice, however, preparing experimentally a perfect density matching solution is challenging. 
Density matching is typically achieved by combining solvents with different mass densities in the correct proportions to match the mass density of the particles~\cite{Kodger2015,Guo2008,Royall2013,Poon2002}.
To sterically stabilize the colloidal particles, they are frequently coated with a polymer layer of a different density than that of the particle core~\cite{Paulin1990,Pusey1986,Ackerson1999,Kooij2000,Kooij2001}.
Due to the polydisperse nature of most colloidal systems, the effective particle density (including both the core and the coating layer) can vary between the particles.
As a result, not all the particles in the solution can have neutral buoyancy.
The buoyant mass of the particles falls within a range roughly centered around neutral buoyancy.
In general there will be particles that have either slightly positive or slightly negative buoyant masses.
We will see here that small deviations from density matching can have a strong effect on sedimentation-diffusion-equilibrium experiments.

\begin{figure}
  \centering
  \includegraphics[width=0.90\linewidth]{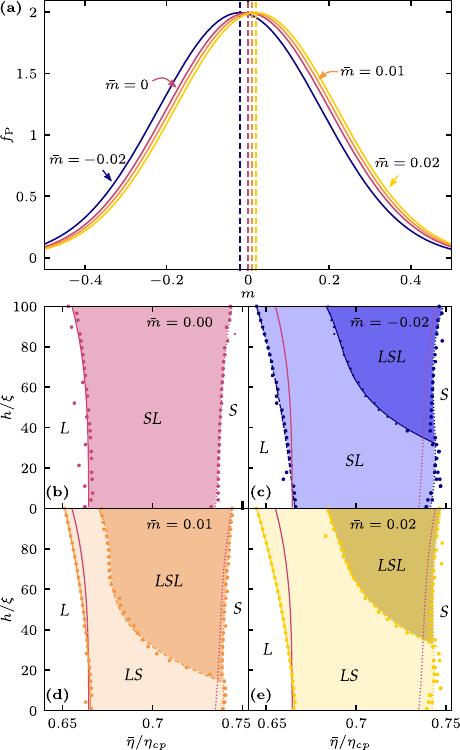}
	\caption{(a) Four parent $\fp$ Gaussian distributions with slightly shifted mean $\mb$ in the range $0\pm 0.02$, as indicated. The standard deviation is $0.2$ in all cases, and the distributions are cut at $m=\pm0.95$ around their respective mean.
	The corresponding stacking diagrams for a mass-polydisperse system of hard spheres in the plane of average packing fraction $\etab$ (relative to close packing $\ecp$) and sample height $h$ (relative to the gravitational length $\xi$) are shown in (b) for the parent distributions with $\mb = 0$, in (c) for $\mb=-0.02$, in (d) for $\mb=0.01$, and in (e) for $\mb=0.02$.
	The sedimentation binodals of paths that end, start, and are tangent to the bulk binodals are indicated with solid-, dashed- and dotted-lines, respectively.
	The symbols are the data points.
	The sedimentation binodals of the case $\mb=0$ are shown for reference in all the stacking stacking  diagrams.
	Note that for the case $\mb=0$ the sedimentation binodals of paths that either start or end at the bulk transition coincide since the parent distribution is symmetrical around $m=0$.
	The bulk system exhibits liquid ($L$) and solid crystalline ($S$) phases.
	The stacking sequences are labeled from the top to the bottom of the sample.
  }
  \label{fig5} 
\end{figure}

We model a system close to density matching by a parent Gaussian distribution $\fp(m)$ roughly centered around a buoyant mass $m=0$, as shown in~\cref{fig5}(a).
We study four different cases, with the mean of the Gaussian $\mb$ sightly shifted in the range of $\pm 0.02$, which is approximately $10 \%$ of their standard deviations.
The distributions are cut at $m=\pm0.95$ around their respective mean.

The stacking diagram for the case $\mb=0$ is shown in~\cref{fig5}(b).
Near density matching, the sedimentation paths are rather horizontal and sensitive to the precise form of the parent distribution.
Hence, the small deviations between the (imposed) target and the (actual) numerical parent distributions that arise in the iterative procedure due to numerical inaccuracies, can have a noticeable
effect.
This is the reason behind the scattered data points (symbols) in the sedimentation binodals of Fig.~\ref{fig5}(b).
With a symmetrical parent distribution around $m=0$ (\ie $\mb=0$) neither particles with positive nor with negative buoyant mass are favored.
Thus, only symmetric stacking sequences (with respect to the midpoint of the sample $z=h/2$) occur, namely $L$, $S$ and $SLS$.
Asymmetric sequences like $LS$ or $SL$ do not appear.

The situation is different for $\mb=-0.02$, where particles with negative buoyant mass that cream up are predominant, see Fig.~\ref{fig5}(c).
Consequently, we also observe the stacking sequence $SL$, with the denser, solid phase, on top of the sample.

In~\cref{fig5}(d) and~\cref{fig5}(e) we show the stacking diagram for the remaining cases $\mb = 0.01$ and $0.02$, respectively. 
For comparison we show always the sedimentation binodals of the buoyant neutral suspension with $\mb = 0$.
The position of the sedimentation binodals for the cases $\mb=-0.02$ and $0.02$ are identical, but the associated stacking sequences are inverted.
This was expected, since changing from $\mb=-0.02$ to $0.02$ is equivalent to inverting the direction of gravity and thus interchanging the meaning of top and bottom of the sample.
This is also the reason why we observe the stacking sequence $LS$ in~\cref{fig5}(e) in the region occupied by $SL$ in~\cref{fig5}(c).
The case $\mb=0.01$ shows the same characteristics as $\mb=0.02$, but with the position of the sedimentation binodals roughly rescaled in the $h/\xi$ axis by a factor of $1/2$, which is the ratio between the mean of the corresponding parent distributions.
Most notable, there is a qualitative difference between the case $\mb=0$ and any other parent distribution considered, namely the lack of asymmetric stacking sequences such as $LS$ and $SL$.
Mass-polydispersity plays therefore an important role in colloidal suspensions close to density matching and even small deviation from density matching can have drastic effects on the stacking diagram.

\subsection{Mass-polydispersity away from density matching}

Not all types of parent distributions are as sensitive to mass-polydispersity as those representing a system near density matching.
In many cases the stacking diagram is robust against perturbations of the parent distribution.
To show this, we construct here four classes of parent distributions and calculate the corresponding sedimentation paths.
The sedimentation paths are quite similar within each class.
We hence can conclude that the corresponding stacking diagrams are also alike.
Recall that the stacking diagram is constructed from the set of special paths, $\mue(z)$, that either start at, end at, or are tangent to the bulk binodal (see~\cref{fig3}).
Thus, if two systems share similar paths for a range of packing fractions and sample heights, then the stacking diagrams will also be similar.

\begin{figure}
  \centering
  \includegraphics[width=\linewidth]{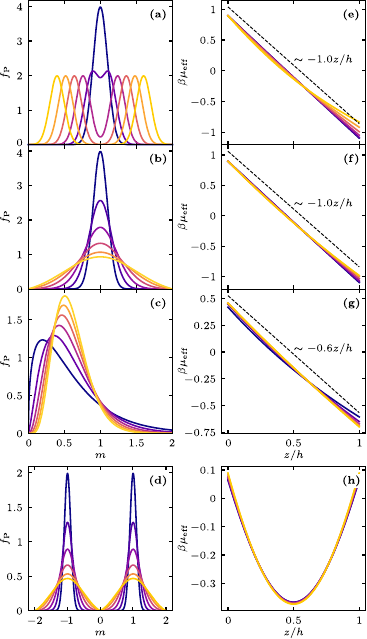}
	\caption{Families of parent distributions $\fp(m)$ in the form of (a) the sum of two Gaussians which move apart symmetrically around the buoyant mass $m=1$, (b) Gaussian with increasing standard deviation from $0.1$ to $0.5$, (c) $\chi^2$-distribution with the degree of freedom increasing from $3$ to $13$, and (d) sum of two Gaussians with mean values at $m=-0.5$ and $m=1.5$ and increasing standard deviation from $0.1$ to $0.5$.
    The distributions in (c) have mean $0.6$ and standard deviation $1$.
    The effective sedimentation paths $\beta\mue$ as a function of the scaled elevation $z/h$ corresponding to the families of distributions in panels (a) to (d) are shown in panels (e) to (h), respectively.
    In all cases we use the Hall EOS for hard spheres, the packing fraction is $\etab/\etacp=0.5$, and the sample height is $h/\xi=80$.
    The dashed lines in panels (e) to (g) are the linear trends of the corresponding sedimentation paths (displaced vertically for a better visualization).
    The slopes (indicated next to each dashed line) coincide in all three cases with the mean mass of the corresponding family of distributions, which are $1$ (a), $1$ (b) and $0.6$ (c).
  }
  \label{fig6}
\end{figure}

The four classes of parent distributions and the corresponding sedimentation paths are shown in~\cref{fig6}.
We construct several distributions within each class by varying a control parameter.
In~\cref{fig6}(a) we increase the mass-polydispersity by interpolating between unimodal and bimodal Gaussian distributions.
In~\cref{fig6}(b) we increase the variance of a Gaussian distribution.
In~\cref{fig6}(c) we vary the skewness of the distribution while keeping the first and the second moment unaltered.
In all cases the distributions contain only positive masses and varying the control parameter has little effect on the sedimentation paths, even when we e.g. drastically increase the degree of mass-polydispersity (second moment of the distribution).
The corresponding sedimentation paths, shown in~\cref{fig6} panels (e) to (g), deviate only slightly from a straight line with a slope given by the mean buoyant mass of the distribution.
Hence, in sedimentation-diffusion-equilibrium, mass-polydisperse systems in which only positive or negative buoyant masses are present are similar to a reference monodisperse system.
(Recall that $\mu(z)=\mu_0-\mbuoy gz$ for a monodisperse system.)
The monodisperse reference system has the same particle mass as the mean of the mass distribution of the mass-polydisperse system.

We also consider a class of parent distributions with both positive and negative buoyant masses, where we increase the variance of a bimodal distribution, see~\cref{fig6}(d).
Due to the presence of buoyant masses with different sign, the suspension does not behave like a monodisperse system under gravity, and hence $\mue(z)$ is not close to a straight line, see~\cref{fig6}(h).
Still the increase in the degree of mass-polydispersity does not affect the behaviour of the system strongly since the paths do not deviate much from each other.

\section{Summary and Conclusions}

Sedimentation path theory~\cite{Heras2013,Geigenfeind2016} was initially developed to study sedimentation-diffusion-equilibrium of binary mixtures.
The theory describes systems that are in equilibrium under the presence of a gravitational field and therefore cannot be used to
describe non-equilibrium phenomena such as drying~\cite{Fortini2016,Kundu2022} or systems that get arrested due to \eg the formation of glasses~\cite{Pusey1986} and non-equilibrium gels~\cite{Harich2016}.
Depending, among other factors, on the the buoyant mass of the colloids, the experimental equilibration times can vary from a few hours to several months~\cite{Heras2012}.
We have extended here sedimentation path theory to deal with mass-polydisperse colloidal systems, \ie the particles are identical except for the value of their buoyant masses.
We derived an exact equation for the sedimentation path of the mass-polydisperse system,~\cref{eq:mu-eff}, that combines all the sedimentation paths of the individual species.
The resulting equation has the structure of the \texttt{LogSumExp} function, often used in machine learning algorithms for its smooth approximation to the maximum function~\cite{Nielsen2016}.
Adding mass polydispersity to a binary mixture is in principle a straight forward extension of the present work.

In bulk, mass-polydispersity has no effect on the phase behaviour.
Hence, our mass-polydisperse model allows us to highlight the interplay between polydispersity and gravity, eliminating by construction the complex effects that shape- and size-polydispersity generates in bulk~\cite{Pusey1987,Bates1998,Auer2001,Sollich2001,MartinezRaton2002,Fasolo2005,Sun2009,Liddle2011,Sollich2011,doi:10.1073/pnas.1909357117}.
Beyond its fundamental interest, a mass-polydisperse system can be specifically realized experimentally by \eg synthesizing core-shell nanoparticles~\cite{Toshima2005,GhoshChaudhuri2012,Wei2011,PajorSwierzy2022} with the same overall size but different relative size between the core and the shell.
In addition, if the degree of size-polydispersity is small, then mass-polydispersity is likely the dominant effect in sedimentation-diffusion-equilibrium.
This is particularly relevant in colloidal suspensions near density matching~\cite{Kodger2015,Guo2008,Royall2013,Poon2002}, in which mass-polydispersity has a big effect on the stacking diagram under gravity: two mass distributions that are only slightly different can give rise to topologically different stacking diagrams containing different stacking sequences.

Granular media is a related system in which the particles can be polydisperse~\cite{PhysRevE.76.021301,PhysRevLett.102.178001,PhysRevE.85.011301,PhysRevE.90.012202}.
It would be interesting to analyse the effects of mass-polydispersity in granular systems. 
For example, phase separation induced by mass-polydispersity might occur in vibrated monolayers~\cite{Narayan2006,GonzalezPinto2017a} of granular systems.

Despite the relevance of the hard-sphere model in soft matter, there are only a few experiments on the sedimentation-diffusion-equilibrium of (quasi) hard-spheres.
Moreover, colloids with a relatively large buoyant mass are often used~\cite{Paulin1990,Ackerson1999} and the sample height is not used as control parameter.
A systematic experimental study of the stacking diagram of hard spheres would be valuable.

In bulk, it is sometimes possible to approximate the free energy of a polydisperse system using only a finite number of moments of the parent distribution~\cite{PhysRevLett.80.1365,moment}.
In a similar way, using the first moment of the parent distribution it is possible to obtain a reasonable approximation for the effective sedimentation path of the mass-polydisperse system, and hence an approximated stacking diagram. 

Polydispersity in the size of the particles affects the bulk behaviour of the suspension and therefore also the sedimentation-diffusion-equilibrium.
For example, \citeauthor{Kooij2001a} studied the sedimentation of polydisperse colloidal platelets~\cite{Kooij2001a}.
Their particle distribution contained platelets of different sizes but only positive buoyant masses.
By changing the overall packing fraction, they found a striking inversion of the stacking sequence from the expected top isotropic and bottom nematic, $IN$, to top nematic and bottom isotropic, $NI$.
Due to the geometric properties of the sedimentation path of a mass-polydisperse system, such inversion of the sequence cannot occur in a mass-polydisperse system that contains particles with only positive (or only negative) buoyant masses.
As correctly pointed out in Ref.~\cite{Kooij2001a}, the inversion must therefore be a consequence of the interplay between gravity and size-polydispersity.
Sedimentation path theory could be applied on top of a bulk theory for size-polydisperse systems in order to describe such effects.

\section*{Data availability}
The data that supports the findings of this study are available within the article 

\begin{acknowledgments}
	We acknowledge useful discussions with Peter Sollich and Stefan Egelhaaf.
  This work is supported by the German Research Foundation (DFG) via project number 436306241.
\end{acknowledgments}

\appendix
\section{Euler-Lagrange equation}
\label{sec:A1}
Carrying out the functional derivative with respect to $\rhom$ in~\cref{eq:Omega-min} for the ideal free energy contribution to $\Omega$ yields
\begin{align}
  \label{eq:Fid-diff}
	\funcd{\Fid{\rhom}}{\rho_{m'}} &= \kT \ln(\rhom).
\end{align}
For the excess contribution we find 
\begin{align}
  \label{eq:Fex-diff-1}
	\funcd{\Fex{\rho}}{\rho_{m'}} = \funcd{\Fex{\rho}}{\rho},
\end{align}
where we have used the functional chain-rule and also the definition of the overall density, Eq.~\eqref{eq:rho}, to calculate the functional derivative
\begin{equation}
	\funcd{\rho}{\rhom} = \funcd{}{\rhom} \intd{m'} \rho_{m'} = \intd{m'}\delta(m-m')= 1.
\end{equation}
Hence, introducing the excess chemical potential $\mu\ind{exc}=\delta\Fex{\rho}/\delta\rho$ in Eq.~\eqref{eq:Fex-diff-1} it follows
\begin{align}
  \label{eq:Fex-diff}
	\funcd{\Fex{\rhom}}{\rho_{m'}} = \mu\ind{exc} = \mu - \kT \ln(\rho).
\end{align}
Here $\mu = \mu\ind{exc} + \kT \ln(\rho)$ is the total chemical potential, including the ideal contribution $\kT \ln(\rho)$, of the corresponding monodisperse system with the same overall density $\rho$ as the mass-polydisperse system.

For the last contribution to $\Omega[\rhom]$ in~\cref{eq:Omega} we get 
\begin{equation}
  \label{eq:mu-diff}
  \funcd{}{\rhom}\intd{m'} \rho_{m'} \mu_{m'} = \mum.
\end{equation}

Hence, adding~\cref{eq:Fid-diff,eq:Fex-diff,eq:mu-diff} according to the minimization principle, Eq.~\eqref{eq:Omega-min}, yields the Euler-Lagrange equation shown in Eq.~\eqref{eq:EL}.

\end{document}